\documentclass[12pt,letterpaper]{imsart}

\RequirePackage{amsthm,amsmath,amsfonts,amssymb,bm}
\RequirePackage[authoryear]{natbib}
\RequirePackage[colorlinks,citecolor=blue,urlcolor=blue]{hyperref}
\RequirePackage{graphicx}% uncomment this for including figures
\usepackage{subcaption}
\usepackage{mwe}

\startlocaldefs

\renewcommand{\Pr}{\mathsf{P}}

\def\by{\bm y}

\def\blambda{\bm \lambda}
\def\balpha{\bm \alpha}
\def\bbeta{\bm \beta}
\def\bmu{\bm \mu}

\def\Pois{\text{Pois}}
\def\Bin{\text{Bin}}

\def\tot{\lambda_\text{tot}}

\endlocaldefs

\begin{document}

\begin{frontmatter}
%%%%%%%%%%%%%%%%%%%%%%%%%%%%%%%%%%%%%%%%%%%%%%
%%                                          %%
%% Enter the title of your article here     %%
%%                                          %%
%%%%%%%%%%%%%%%%%%%%%%%%%%%%%%%%%%%%%%%%%%%%%%
\title{Multi-scale Poisson process approaches for differential expression analysis of high-throughput sequencing data}
%\title{A sample article title with some additional note\thanksref{T1}}
\runtitle{multiseq}
%\thankstext{T1}{A sample of additional note to the title.}

\begin{aug}
%%%%%%%%%%%%%%%%%%%%%%%%%%%%%%%%%%%%%%%%%%%%%%
%%Only one address is permitted per author. %%
%%Only division, organization and e-mail is %%
%%included in the address.                  %%
%%Additional information can be included in %%
%%the Acknowledgments section if necessary. %%
%%%%%%%%%%%%%%%%%%%%%%%%%%%%%%%%%%%%%%%%%%%%%%
\author[A]{\fnms{Heejung} \snm{Shim}\ead[label=e1]{heejung.shim@unimelb.edu.au}},
\author[B]{\fnms{Zhengrong} \snm{Xing}\ead[label=e2,mark]{xiamanoobix@gmail.com}},
\author[B]{\fnms{Ester} \snm{Pantaleo}\ead[label=e3,mark]{esterpantaleo@gmail.com}},
\author[C]{\fnms{Francesca} \snm{Luca}\ead[label=e4]{fluca@wayne.edu}},
\author[D]{\fnms{Roger} \snm{Pique-Regi}\ead[label=e5]{rpique@wayne.edu}}
\and
\author[E]{\fnms{Matthew} \snm{Stephens}\ead[label=e6]{mstephens@uchicago.edu}}
%%%%%%%%%%%%%%%%%%%%%%%%%%%%%%%%%%%%%%%%%%%%%%
%% Addresses                                %%
%%%%%%%%%%%%%%%%%%%%%%%%%%%%%%%%%%%%%%%%%%%%%%
\address[A]{School of Mathematics and Statistics and Melbourne Integrative Genomics, University of Melbourne, Melbourne, 3010, Australia, \printead{e1}}

\address[B]{Department of Statistics, University of Chicago, Chicago, IL 60637, USA, \printead{e2,e3}}

\address[C]{Department of Obstetrics and Gynecology and Center for Molecular Medicine and Genetics, Wayne State University, Detroit, MI 48201, USA, \printead{e4}}

\address[D]{Center for Molecular Medicine and Genetics, Wayne State University, Detroit, MI 48201, USA, \printead{e5}}

\address[E]{Department of Statistics and Human Genetics, University of Chicago, Chicago, IL 60637, USA, \printead{e6}} 

\end{aug}

\begin{abstract}
Estimating and testing for differences in molecular phenotypes (e.g.~gene expression,  chromatin accessibility, 
%methylation
transcription factor binding
) across conditions is an important part of understanding the molecular basis of gene regulation. These phenotypes are commonly measured using high-throughput sequencing assays (e.g., RNA-seq, ATAC-seq, ChIP-seq), which provide high-resolution count data that reflect how the phenotypes vary along the genome. Multiple methods have been proposed to help exploit these high-resolution measurements for differential expression analysis. However, they ignore the count nature of the data, instead using normal approximations that work well only for data with large sample sizes or high counts. Here we develop count-based methods to address this problem. We model the data for each sample using an inhomogeneous Poisson process with spatially structured underlying intensity function, and then, building on multi-scale models for the Poisson process, estimate and test for differences in the underlying intensity function across samples (or groups of samples). Using both simulation and real ATAC-seq data we show that our method outperforms previous normal-based methods, especially in situations with small sample sizes or low counts. 
\end{abstract}


\end{frontmatter}
%%%%%%%%%%%%%%%%%%%%%%%%%%%%%%%%%%%%%%%%%%%%%%
%% Please use \tableofcontents for articles %%
%% with 50 pages and more                   %%
%%%%%%%%%%%%%%%%%%%%%%%%%%%%%%%%%%%%%%%%%%%%%%
%\tableofcontents

%%%%%%%%%%%%%%%%%%%%%%%%%%%%%%%%%%%%%%%%%%%%%%
%%%% Main text entry area:

\section{Introduction}

%\textcolor{red}{[HJ's comments on Introduction: 1) Li Ma's method "Analysis of Distributional Variation Through Graphical Multi-Scale Beta-Binomial Models. https://www.tandfonline.com/doi/full/10.1080/10618600.2017.1402774"  This method detects differences in DNase-seq data in their application. But my understanding is they treats read counts along the region as a histogram representing a distribution, and applied their method which detect a difference in distributions. It means they ignore difference in total counts (my understanding could be wrong). If my understanding is correct, their application is quite misleading. I'm not sure if we should cite this paper; 2) I cited \cite{lee2016} "Identification of differentially methylated loci using wavelet-based functional mixed models" by Wonyul Lee and Jeffrey S. Morris https://pubmed.ncbi.nlm.nih.gov/26559505/ as previous methods that better use the high-resolution measurements to detect differences in molecular phenotypes. However, the application is somewhat different from the context we consider. The sentence from the paper "Contrary to many commonly-used methods that model probes independently, this framework accommodates spatial correlations across the genome through basis function modeling". But I think its ok to cite this paper as a previous method..]}

To understand the molecular basis of gene regulation,
scientists often study the way that molecular phenotypes---such as gene expression \citep{Marioni_2008, moyerbrailean2015}, chromatin accessibility \citep{buenrostro2013},  and transcription factor binding \citep{luca2013, buenrostro2013} ---vary among samples or treatment conditions (e.g., different environments, developmental stages, or tissues). These phenotypes
are usually measured using sequencing-based assays---such as
RNA-seq \citep{Mortazavi_2008a, Wang_2008, Marioni_2008}, ChIP-seq \citep{Johnson_2007, Barski_2007, Mikkelsen_2007}, DNase-seq \citep{Boyle_2008, Hesselberth_2009}, and ATAC-seq \citep{buenrostro2013}---which are cheap and high throughput. Collectively we refer to these assays
as $\ast$-seq. These $\ast$-seq assays provide high-resolution measurements across the whole genome that reflect how the molecular phenotypes vary along the genome. Specifically, 
for each location in the genome, these assays provide a
``count'' of the number of sequences that arose from that location, where the size of the count reflects the intensity of the underlying phenotype at that location. 
In this paper we develop methods to detect and estimate {\it differences} in this intensity among samples along the genome,
taking account of both the high-resolution and the count nature of the data.

%For example, the HTS data are commonly summarized by the number of ``reads" mapping to each base in the genome. 
% For example, in ATAC-seq data for each sample can be summarized by the counts of ``reads'' starting at each base in the genome that approximately represent chromatin accessibility (i.e., open chromatin) at the base. 

%Most widely-used differential analysis methods for the HTS data (e.g., edgeR \cite{robinson2010edger}, DESeq \cite{anders2010DESeq}, DESeq2 \cite{love2014DESeq2}, limma+voom \cite{law2014voom}) have been developed for differential gene expression analysis of RNA-seq data, and aimed to detect differences in ``overall expression" of genes. Those ``overall expression" methods combine the data within genes (or predefined regions) to quantify overall expression, and do not exploit the high-resolution measurements in the HTS that capture ``shape" of expression within genes. On the other hand, a method introduced by \cite{ma2016} can be used to identify differences in ``shape" of the phenotypes. 

The simplest approach to this problem is to 
divide the genome into regions, add up the counts in each sample in each region, and then test for differences in these total counts using one of a wide range of analysis methods available for this task (e.g., DESeq2 \citep{love2014DESeq2}, edgeR \citep{robinson2010edger}, limma+voom \citep{law2014voom}). We refer to these methods as
``overall expression methods'' because their most common
usage is in gene expression studies, where the regions usually correspond to genes, and the methods aim to detect differences in the overall expression of each gene. The main limitation
of this approach is the difficulty of selecting an appropriate region size: if regions are too small then the methods will have no power due to low counts; if regions are too big then one loses the sensitivity of the inference, and indeed one risks missing signals that affect smaller subregions. In other words these approaches, while simple, do not 
fully exploit the high-resolution measurements in $\ast$-seq data.

%To address this problem several authors
%have developed methods that aim to make better use
%of the high-resolution measurements
%\citet{Frazee2014, torres2016derfinder, Shim2015, lee2016}. For example, \citet{Shim2015} proposed a wavelet-based method, WaveQTL, which tests for differences at multiple resolutions simultaneously, combining information across resolutions to avoid the problems of selecting a single resolution or region size.
%\textcolor{red}{Other methods....} 
To address this problem several authors have developed methods that aim to make better use of the high-resolution measurements \citep{Shim2015, lee2016, Frazee2014, torres2016derfinder}. \citet{Shim2015} and \citet{lee2016} proposed wavelet-based methods which test for differences at multiple resolutions simultaneously, combining information across resolutions to avoid the problems of selecting a single resolution or region size. \citet{Frazee2014} and \citet{torres2016derfinder} presented approaches which compute base-resolution test statistics along the genome, identify regions of consecutive bases showing a common differential expression signature, and test for each region by combining information across bases in the region. The main limitation of these methods is that they use a normal distribution to model the read counts, which can work well if counts are sufficiently high, or sample sizes are sufficiently large, but performs poorly for small sample sizes or low counts \citep{Shim2015}. Finally, \citet{ma2018} introduced a multi-scale method for analysis of distributional variance which uses hierarchical nonparametric models. However, this method focuses on modelling differences in relative frequency at multiple resolutions, ignoring differences in overall expression in regions, so it is not directly applicable to typical differential expression analysis. 

%In this paper, we introduce a method, multiseq, to address this problem by directly modelling the count nature of the data (see Supplementary Material Table 1). 
In this paper, we introduce a method, multiseq, that 1) better exploits high-resolution measurements in the $\ast$-seq data, as well as 2) directly models the count nature of the data (see Supplementary Material Table 1). 
Specifically, we assume that the count data for each sample are generated by an inhomogeneous Poisson process with a spatially structured underlying intensity function, and estimate this intensity function using extensions of existing multi-scale models for inhomogeneous Poisson processes \citep{Kolaczyk_1999, Timmermann_1999,xing.stephens.2016}. A key innovation 
of our method is that it estimates and tests for the {\it differences} in the underlying intensity among samples, taking account of the fact that these difference will be spatially structured.
We illustrate the benefits of multiseq--- particularly for small sample sizes or low counts---using both simulation and real data analyses. We apply mutliseq to ATAC-seq data of 3 copper treated and 3 control samples for a large-scale differential chromatin accessibility analysis involving hundreds of thousands of tests. We find that multiseq identifies 1083 differences at the false discovery rate 0.05, which is 2.6 and 1.6 times the number of differences identified by a normal-based method and a simpler overall expression method, respectively. Our methods are implemented in the R package multiseq, available at \url{https://github.com/heejungshim/multiseq}.
%We illustrate the benefits of multiseq--- particularly
%for small sample sizes or low counts---using both simulation
%and real data analyses. Our methods are implemented in the R package multiseq, available at \url{https://github.com/heejungshim/multiseq}.

%The coarsest scale in multi-scale model captures ``overall" expression of phenotypes in regions. Thus, multiseq can be considered as an extension of widely-used ``overall expression methods" for differential analysis, such as DESeq2 \citep{love2014DESeq2} and edgeR \citep{robinson2010edger}, that directly model the total number of reads mapping to the regions as count data. In addition to the total read counts, multiseq exploits high-resolution measurements in the data through the remaining scales in the model. Indeed, our software provides an option to incorporate outputs from the overall expression methods into multiseq at the coarsest scale. 

% Here, we present a novel method, multiseq, for identifying and estimating the differences in the molecular phenotypes between multiple groups using $\ast$-seq data.

\section{Materials and methods}

% \textcolor{red}{Matthew's comments on this section: I'm going to assume this is the "Detailed Methods" section and write it as such. We will also want a less technical methods overview later.}

Assume that we have $\ast$-seq data from $n$ samples across a genomic region divided into $B$ equal bins. Specifically, the data consist of
the ``read counts'' in each sample that map to each bin. Here, ``map to a bin'' means that the {\it start} of the read falls in the bin, so each read maps to exactly one bin.
(The choice of bin-width constitutes a trade-off between computation and
resolution. For example, the highest possible resolution of analysis would be at the single base level: each bin would be of length 1bp. However, computation could
be reduced -- typically by a factor of $>$10 -- by using bins of length 10bp.)
Let $y^{i}_b$ denote the read count for sample $i$ in bin $b$, and let $\by^{i}$ denote the vector $(y_1^{i}, \dots, y_B^{i})$.

We model the read counts as arising from an inhomogeneous Poisson process:
\begin{equation} \label{eq:pois}
y^i_b \sim \Pois(\lambda^i_b).
\end{equation}
At a high level our goal is to estimate $\lambda^{i}_b$,
and in particular to identify which bins $b$ show differences in $\lambda^{i}_b$ among samples $i$ (or between groups of samples). 
%Our methods are designed to address two key challenges. First, 
Our methods are designed to address the following challenge. The information available for each bin $b$ is very limited. For example, in the extreme case where each bin contains a single base, the typical number of reads in each bin could be 0 or 1. In this case, analyzing the data bin-by-bin will be hopeless. One solution would
be to make the bins big enough to make such bin-by-bin analysis tractable. However, if the bins are made too big, this approach risks missing fine-resolution signals. 
To avoid this, we instead take a ``multi-scale'' approach to combine information across bins, exploiting the idea that nearby bins will often (though not always) tend to have
similar effects, to smooth estimates across bins. 
%Second, our methods aim to address the fact that we often expect most bins $b$ to be ``null": that is, for many bins $b$ $\lambda^i_b$ will be similar across samples $i$.

To implement this in practice we generalize previous multi-scale smoothing methods for inhomogeneous Poisson processes \citep{Kolaczyk_1999, Timmermann_1999, xing.stephens.2016}. These methods essentially fit \eqref{eq:pois} for a {\it single} sample $i$, combining information across $b$. We review these methods in the next
subsection before describing our extension to multiple samples.

%\textcolor{red}{HJ: this section can go to Introduction.} In the proposed multi-scale approach, a model for association of $g$ with a total read count over the site (i.e., $\sum_{b=1}^{B} y_b^i$) is equvalent \textcolor{red}{(very similar?)} to models considred by traditional approaches which use only the total read counts for differential expression analysis (e.g., edgeR \cite{robinson2010edger}, DESe \cite{anders2010DESeq}, DESeq2 \cite{love2014DESeq2}). That is, our approach is an extension of the traditional approaches to use high-resolution measurement (`shape' information) in ATAC-seq data in addition to the total read counts.

\subsection{Multi-scale models for inhomogeneous Poisson processes: single sample}

%Here, we applied the proposed methods to ATAC-seq data \citep{buenrostro2013} for differential chromatin accessibility analysis, but multiseq could be applied to differential analysis of other $\ast$-seq data.
Consider the model \eqref{eq:pois} for a single sample $i=1$.
To lighten notation we drop the superscript $i$, 
and so the observed data are $\by =(y_1,\dots,y_B)$ with
\begin{equation} \label{eqn:pois_1d}
y_b \sim \Pois(\lambda_b).
\end{equation}
Consider estimating $\blambda=(\lambda_1,\dots,\lambda_B)$ under the assumption that 
 $\blambda$ is ``spatially structured'': that is, where
$|\lambda_b - \lambda_{b+1}|$ is small for most (but not necessarily all) $b$. 
For simplicity we assume that $B$ is a power of 2, so $B = 2^J$ for some $J$.

The multi-scale approach to this problem involves two steps:
\begin{enumerate}
\item Reparameterize the Poisson model \eqref{eqn:pois_1d} using a 1-1
``multi-scale'' transformation \citep{Kolaczyk_1999, Timmermann_1999},
$\blambda = f_{\text{ms}}(\balpha, \tot)$ 
defined below. 
The key property of
this reparameterization is that each element of 
$\balpha$ captures the spatial variation in $\blambda$ at a particular scale (resolution) and location. Consequently one can capture the idea that $\blambda$ is spatially structured by modelling sparsity of $\balpha$, which is relatively straightforward. This is similar to the key idea of wavelets \citep{Donoho1995}, which are widely used in this way for Gaussian processes. 
\item Perform shrinkage-based estimation of $\balpha$. \cite{xing.stephens.2016} do this by
making a normal approximation to the likelihood for $\balpha$, and then using the Empirical Bayes (EB) shrinkage methods from \cite{stephens2016}.
\end{enumerate}
Here we briefly summarize each step; see \cite{xing.stephens.2016} for further details.
%Other applications of multi-scale models include \citet{raj2015} which used the idea for inference of transcription factor binding sites. 

%\def\by{\bm y}
%\def\blambda{\bm \lambda}
%\def\balpha{\bm \alpha}
%\def\bbeta{\bm \beta}
%\def\bmu{\bm \mu}
%\def\bp{\bm p}
%\def\Poi{\text{Poi}}
%\def\Bin{\text{Bin}}
%\def\BF{\text{BF}}
%\def\tot{\lambda_\text{tot}}

At scale $s=1,\dots,\log_2(B)$ define $2^{s-1}$ ``locations'' $l$ by dividing the indices $\{1,\dots,B\}$ into $2^{s-1}$ equal groups of consecutive indices, and let $I_{sl}$ denote the indices of location $l$ at scale $s$ so formed.
For example, at scale $3$ there are $2^2 = 4$ locations, and: 
\begin{align}
    I_{31} &= [1,B/4]; \\
    I_{32} &= [B/4+1,B/2]; \\
    I_{33} &= [B/2 +1,3B/4]; \\
    I_{34} &= [3B/4+1,B].
\end{align}
where $[a,b]$ denotes the indices $(a,a+1,\dots,b)$.
Further let $I^{-}_{sl},I^{+}_{sl}$ respectively denote the first and second halves of the indices in $I_{sl}$. So, for example, $I^-_{31} = [1,B/8]$ and $I^+_{31} = [B/8+1,B/4]$.

Now let $\lambda^-_{sl}, \lambda^+_{sl}$ respectively denote the sums of the values of $\lambda$ across indices $I^-_{sl}$ and $I^+_{sl}$.
So,
\begin{equation}
    \lambda^-_{sl} := \sum_{i \in I^-_{sl}} \lambda_i; \qquad 
    \lambda^+_{sl} := \sum_{i \in I^+_{sl}} \lambda_i. 
\end{equation}
Finally define the multi-scale parameters $\balpha$ by
\begin{equation}
    \alpha_{sl} := \log [\lambda^-_{sl}/\lambda^+_{sl}].
\end{equation}
The intuition is that $\alpha_{sl}$ 
captures the change in (log) intensity 
between the first and second half of location $I_{sl}$. In particular, if
$\lambda$ is constant
in $I_{sl}$ then $\alpha_{sl}=0$.
When combined with the total intensity, $\tot:=\sum_{i=1}^b \lambda_i$, the $\balpha$ represent a 1-1 ``multi-scale'' reparameterization of $\blambda$, and we write $\blambda = f_{\text{ms}}(\balpha, \tot)$.

A key property of this reparameterization
is that the likelihood factorizes into independent terms \citep{Kolaczyk_1999}. Specifically:
\begin{equation} \label{eq:fact}
    p(\by; \balpha, \tot) = \Pois(\sum_b y_b; \tot) \prod_{sl} \Bin \left(y_{sl}^{-}; y_{sl}^{-}+ y^+_{sl},  \exp(\alpha_{sl})/(1+\exp(\alpha_{sl}) ) \right)
\end{equation}
where $\Pois(\cdot; \lambda)$ denotes the  probability mass function (pmf) of the Poisson distribution with parameter $\lambda$, $\Bin(\cdot; n,p)$ denotes the pmf of the Binomial distribution with $n$ trials and success probability $p$, and $y^-_{sl},y^{+}_{sl}$ denote, respectively, the sum of $y_b$ over the indices $I^-_{sl},I^+_{sl}$. This result
follows from the elementary distributional result:
if $y_1,y_2$ are independent, with $y_j \sim \Pois(\lambda_j)$ then 
\begin{align}
p(y_1, y_2 | \lambda_1,\lambda_2) &= \Pois(y_1+y_2; \lambda_1 + \lambda_2) \Bin(y_1; y_1+y_2, \lambda_1/(\lambda_1+\lambda_2).
\end{align} 

From \eqref{eq:fact} we note that the information in the data about $\alpha_{sl}$ is exactly that contained in a single binomial observation, which can be written
\begin{align} \label{eq:ysl}
    y^-_{sl} &\sim \Bin(y_{sl}^{-}+ y^+_{sl}, p_{sl}) \\
    \alpha_{sl} &= \log(p_{sl}/(1-p_{sl})).
\end{align}

With this reparameterization in place,
the second step of the multi-scale approach is to perform shrinkage estimation of $\balpha$.
\citet{xing.stephens.2016} perform this shrinkage
estimation by introducing normal approximations to the binomial likelihoods in \eqref{eq:fact}, 
giving an (approximate) likelihood $L(\balpha; \by)$ of the form:
\begin{equation} \label{eq:lik_alpha_approx}
L(\balpha; \by) = \prod_{sl} N(\hat{\alpha}_{sl}; \alpha_{sl}, s^2_{sl}), 
\end{equation}
where $N(\cdot; \mu, \sigma^2)$ denotes the density of a normal distribution with mean $\mu$ and variance $\sigma^2$, 
and $\hat{\alpha}_{sl}, s^2_{sl}$ are respectively estimates for $\alpha_{sl}$ and its standard error based on a binomial observation. 
This allows them to 
exploit the flexible EB shrinkage methods from \cite{stephens2016} to perform the shrinkage
estimation.
Specifically these methods perform
shrinkage by estimating a prior distribution, $g$,
for the elements of $\balpha$ of the form:
\begin{align} \label{eq:mix_normal}
g(\cdot; {\bm \pi}) = \pi_0 \delta_0(\cdot) + \sum_{k=1}^K\pi_k N(\cdot; 0, \sigma_k^2),
\end{align}
where $\delta_0(\cdot)$ denotes a point mass at 0 and $\sigma_1, \ldots, \sigma_K$ dente a large and dense grid of fixed positive numbers spanning a range from very small to very big. The mixture proportions ${\bm \pi} = (\pi_0, \ldots, \pi_K)$, which are non-negative and sum to one, are  estimated by maximum likelihood. The posterior on the elements of $\balpha$
given this estimated prior then has a closed form. See \cite{stephens2016} for details.

\cite{xing.stephens.2016} also describe other details
to improve performance. In particular, to allow for different amounts of shrinkage at different scales, they use scale-specific mixing proportions $\hat{\bm \pi}^s$ in \eqref{eq:mix_normal}, and estimate these using a ``translation-invariant'' transform \citep{coifman.donoho} that 
efficiently averages results over all translations of the data.
This averaging across translations also helps avoid artifactually large changes in estimated $\blambda$ that can otherwise occur (for example) exactly half-way across the region.

\subsection{multiseq: multi-scale models for inhomogeneous Poisson processes from multiple groups of samples}

The main methodological contribution of our paper is to extend the ideas above to the multi-sample case.
Specifically, we consider the model:
\begin{equation}
%y^{i}_b | \quad \balpha^i, \tot^i \sim \Pois(\lambda_b^i) \quad i=1,\dots,n
y^{i}_b | \balpha^i, \tot^i \sim \Pois(\lambda_b^i) \quad i=1,\dots,n,
\end{equation}
where $\lambda_b^i$ represents the $b$-th component of the multi-scale transformation $\blambda^i = f_{\text{ms}}(\balpha^i, \tot^i)$ described above.

We assume that a covariate $X^i$ is measured on each sample $i$; for concreteness we assume $X^i \in \{0,1\}$ is a binary group membership, but our methods
apply equally to a continuous covariate. To model the effect of $X^i$ on the intensity $\blambda^i$, we introduce a linear model for the multi-scale parameters $\balpha^i$:
\begin{equation}
\alpha_{sl}^i = \mu_{sl}+\beta_{sl}X^i+u_{sl}^i, \label{eq:multiscale}
\end{equation}
where $\mu_{sl}$ denotes the mean of $\alpha_{sl}$ for samples with $X^i = 0$, $\beta_{sl}$ denotes the effect of $X^i$ on $\alpha_{sl}$, 
and $u_{sl}^i$ denotes a zero-mean individual-specific random effect to model both over-dispersion and biological variability among samples. 

Our goal now is to perform shrinkage-based estimation
of the multi-scale parameters $\mu_{sl}$, which captures the mean Poisson intensity across samples, and---more importantly for our purposes---$\beta_{sl}$, which captures the difference in intensity between groups. By performing
shrinkage estimation we capture that both the underlying intensity and the difference in intensity between groups are expected to be spatially structured.

As in the single-sample case, the likelihood for the multi-scale parameters factorizes across $s,l$. In particular, analogous to \eqref{eq:ysl}, the information in $\mu_{sl},\beta_{sl}$ is  contained in $n$ binomial observations:
\begin{align} 
y^{i,-}_{sl} & \sim \Bin(y^{i,-}_{sl}+ y^{i,+}_{sl}, p^i_{sl}),  \label{eq:y16}\\
\log(p^i_{sl}/(1-p^i_{sl})) & = \alpha^i_{sl} = \mu_{sl} + \beta_{sl}X^i + u_{sl}^i \qquad i=1,\dots,n \label{eq:y17}.
\end{align}
This is a generalized (binomial) linear mixed model, and so we can use standard methods to obtain
estimates $\hat{\mu}_{sl}, \hat{\beta}_{sl}$ and their corresponding standard errors $s_{\mu_{sl}}, s_{\beta_{sl}}$. Similar to \cite{xing.stephens.2016} we then use these estimates and standard errors to form a normal approximation to the likelihood,
\begin{equation} \label{eq:lik_joint_approx}
L(\bmu, \bbeta; Y) = \prod_{sl} N(\hat\mu_{sl}; \mu_{sl}, s^2_{\mu_{sl}}) N(\hat\beta_{sl}; \beta_{sl}, s^2_{\beta_{sl}}),
\end{equation}
and apply the methods from \citet{stephens2016} to produce shrinkage estimates of $\bmu$ and $\bbeta$.  

This shrinkage estimation approach leads to smooth estimates of both the mean log-intensity in each group and the difference in log-intensity between the two groups. The methods also provide posterior distributions for $\bmu$ and $\bbeta$ and the marginal likelihood integrating out $\bmu$ and $\bbeta$. Details are given in Supplementary Material. In addition, Supplementary Material describes:
\begin{itemize}
    \item A standard approach to fit the generalized linear mixed model provides estimates and standard errors for $\bmu$ and $\bbeta$, but we modified this procedure slightly to improve performance in practice.
    \item The likelihood for $\mu_{sl},\beta_{sl}$ in~\eqref{eq:y16}-\eqref{eq:y17} does not in general factorize into a term for $\mu_{sl}$ and a term for $\beta_{sl}$. Thus, to improve the approximation in~\eqref{eq:lik_joint_approx}, we reparameterize, following \citet{wakefield2009}, from $\mu_{sl}$, $\beta_{sl}$ to $\mu_{sl}^\ast$, $\beta_{sl}$, whose likelihoods asymptotically factorize into two independent terms.
    \item The effect of  $X$ on the overall expression (intensity) $\lambda_\text{tot}$ can be modelled in multiple ways. Our software provides two approaches that are based on generalized linear models with random effects: Poisson regression and binomial regression. Both approaches address the issue of different sequencing depths across samples. The former has been adapted by widely-used overall expression methods for differential analysis (e.g., DESeq2 \citep{love2014DESeq2} and edgeR \citep{robinson2010edger}). Thus, our software provides an option to incorporate outputs from the existing methods for the Poisson regression approach. We used the output from DESeq2 in the analysis of this paper.
\end{itemize}

\subsubsection{Testing for differences in molecular phenotype over the region between multiple groups of samples} To test for non-zero differences (i.e., effects) over the region, we test the null hypothesis $H_0: \beta_{sl}=0$ $\forall s, l$.  This is equivalent to testing $\pi_0 = 1$ in the prior \eqref{eq:mix_normal} for $\beta$, which, following \citet{Shim2015}, we test using the likelihood ratio test statistic
\begin{equation} \label{eq:likeli_statistic}
\Lambda = \prod_{sl} \frac{\Pr(\hat\beta_{sl} | s^2_{\beta_{sl}}, \hat{\bm \pi})}{ \Pr(\hat\beta_{sl} | s^2_{\beta_{sl}}, \pi_0 = 1)}, 
\end{equation} 
where $\hat{\bm \pi}$ denotes the maximum likelihood estimate $\hat{\bm \pi} :=\arg\max\prod_{sl}\Pr(\hat\beta_{sl} | s^2_{\beta_{sl}}, {\bm \pi})$; see Supplementary Material for details of the marginal likelihood $\Pr(\hat\beta_{sl} | s^2_{\beta_{sl}}, {\bm \pi})$. We assess the significance of the likelihood ratio test statistic using its empirical distribution under $H_0$. The empirical distribution can be obtained by permutation (i.e., permutation of sample labels for $X$; see \citet{Shim2015}) or by using a data set that is expected to have no difference in phenotype among samples (e.g., by comparing controls vs controls; see our analysis in section~\ref{sec:results}).

\subsubsection{Effect size estimation} The factorized normal likelihood and mixture priors on $\bbeta$ yield closed forms for the posterior distributions; see Supplementary Material. To provide more interpretable estimates of the effect of $X$, we wish to convert these posteriors in the multi-scale space to posteriors for the underlying log-intensity function $\log{\bm \lambda}$ in the original observation space.
 We define the effect on base $b$ in
the observation space as $\beta^o_b:=\log(\lambda_{b}^{(1)}/\lambda_{b}^{(0)})$, 
where $\blambda^{(0)}, \blambda^{(1)}$ denote the values for $\blambda$
for an individual in group $0$ and $1$ respectively, whose random effects $u$ in \eqref{eq:multiscale} are set to 0.
The posterior on $\bbeta^{o}$ does not have a simple analytic form. However, we can approximate the pointwise posterior mean and variance  using Taylor series approximations (see Supplementary Material). Other types of posterior inference could be performed by sampling from the posterior for $\bbeta^{o}$,
which can be achieved by first simulating samples from the posteriors on $\bmu$ and $\bbeta$, and transforming them to posterior samples for $\bbeta^{o}$ using the relationship with ${\bm \lambda}$ (see Supplementary Material).

\section{Results} \label{sec:results}
In this section, we assess the performance of multiseq against existing methods on both simulated and real data sets. We first illustrate the advantages of multiseq compared to WaveQTL \citep{Shim2015}, which fails to directly model the count nature of $\ast$-seq data, in a simulation study. Then, we show the application of multiseq to real data with small sample size, and compare multiseq with WaveQTL and an overall expression method, DESeq2 \citep{love2014DESeq2}.

\subsection{Simulation study}
To demonstrate the benefits of multiseq in the analysis of data sets with small sample sizes or low read counts, we compared multiseq with WaveQTL on simulated data. (Supplementary Material presents results for the overall expression method DESeq2 \citep{love2014DESeq2}, which performs consistently less well than multiseq here because it does not exploit the high-resolution information in the data.)

Following \citet{Shim2015}, we simulated realistic data by subsampling reads from a real DNase-seq data set from \cite{Degner_2012} which measures chromatin accessibility  \citep{Boyle_2008} along the genome. These data consist of $\approx$ 2.75 B reads for 70 samples ($\approx$ 39 M reads per sample). We performed simulations using the 544 regions of length 1024bp that were reported as associated with genetic variants in \citet{Shim2015}. 
For each region, we simulated two data sets for varying sample sizes (6, 10, and 70) and expected library read depths (39$\times$0.1 M, 39$\times$0.5 M, 39 M, 39$\times$2 M, 39$\times$ 4 M), each of which has two groups with an equal number of samples in each group. One of the two data sets is null (i.e., no effect) and the other is non-null, with the effect given by the estimated true effects from \citet{Shim2015}. See Supplementary Material for the detailed simulation procedure. This procedure results in 1088 data sets (544 null and 544 non-null) for each sample size and expected library depth. 

We applied multiseq and WaveQTL to each simulated data set, obtained the test statistic for each method (the likelihood ratio test statistic in~\eqref{eq:likeli_statistic} for multiseq and a similar likelihood ratio statistic for WaveQTL; see \citet{Shim2015} for details), and assessed the performance of each method based on area under the receiver operating characteristic curve (AUC).

\subsubsection{multiseq has a potential to better maintain power at small sample sizes or for low read counts compared to WaveQTL} 
\begin{figure}[h]
\includegraphics[scale=0.11]{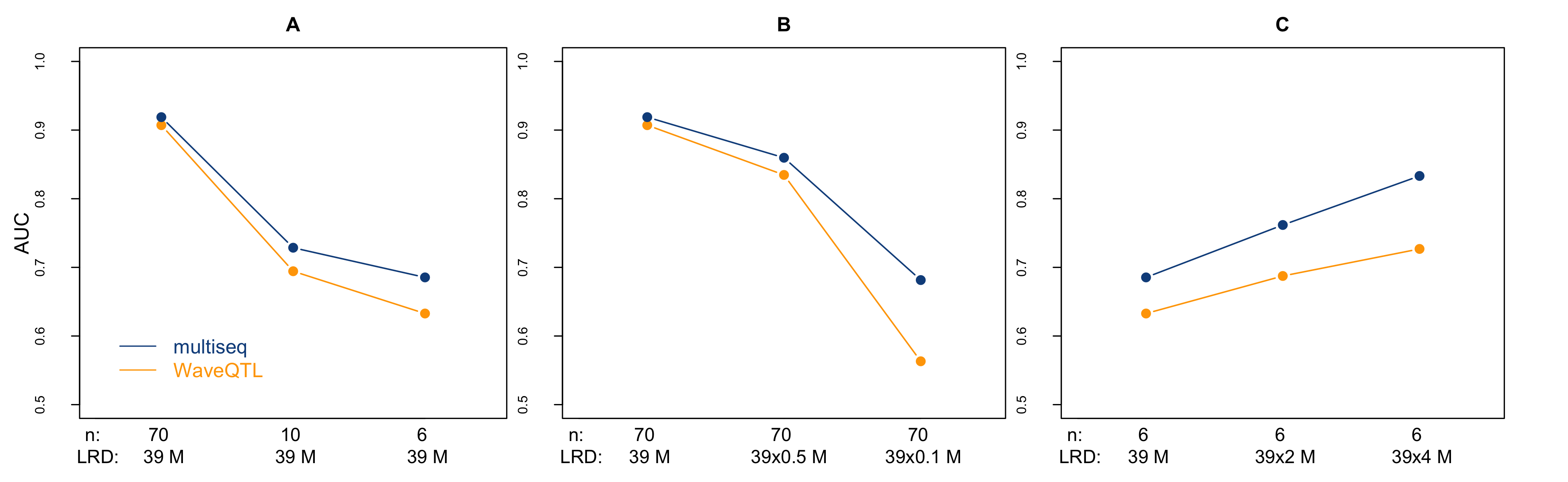}
\caption {{\bf multiseq has a potential to better maintain power at small sample sizes or for low read counts compared to WaveQTL.} Performance of multiseq (in blue) and WaveQTL (in orange) is assessed at different sample sizes ($n$) or expected library read depths (LRD) by using area under the receiver operating characteristic curve (AUC). Panel A shows the performance at the expected library read depth of 39 M as sample size varies over 70, 10, and 6. Panel B shows the performance at sample size 70 as the expected library read depth decreases from 39 M to 39$\times$0.5 M, 39$\times$0.1 M. Panel C shows the performance at the sample size 6 as the expected library read depth increases to 39$\times$2 M and 39$\times$4 M.}
\label{fig:compareAUC}
\end{figure}

We first present results for varying sample sizes (6, 10, and 70) at a sinlge expected library read depth (39 M). Compared with WaveQTL, multiseq achieved higher AUC at all three sample sizes, with larger difference in performance at smaller sample sizes (Fig~\ref{fig:compareAUC}A). 

Next we present results for varying library read depth (39 M to 39$\times$0.5 M, 39$\times$0.1 M) at fixed sample size (70). The results in Fig~\ref{fig:compareAUC}B show that multiseq achieved higher AUC than WaveQTL at all three library read depths. In particular, for the lowest expected read depth of 39$\times$0.1 M, WaveQTL has almost no power to detect signals (AUC=0.56), while multiseq still shows a moderate power (AUC=0.68). 

In summary, these simulations illustrate the potential of multiseq, which directly models the count nature of $\ast$-seq data, to better maintain power at small sample sizes or for low read counts compared to WaveQTL.  

\subsubsection{For small sample sizes, increasing library read depths leads to better performance of multiseq} Previous studies have shown that power of 
overall expression methods to detect effects in lowly-expressed regions can be increased by  increasing library read depth \citep{tarazona2011differential, robinson2014subseq}.  Fig~\ref{fig:compareAUC}B shows that power
of  both multiseq and WaveQTL increases with library read depth at the relatively large sample size 70. Now we assess the performance improvement at the smaller sample size 6. The results (Fig~\ref{fig:compareAUC}C) confirm that increased read depths lead to better performance for multiseq. For example, the AUC for multiseq at the sample size 6 with 39$\times$2 M and 39$\times$4 M read depths (0.76 and 0.83, respectively) are higher than the AUC of multiseq at the larger sample size 10 with 39 M library read depth (0.72). However, consistent with previous studies of overall expression methods \citep{liu2014rna, Scotty2013},
there is an upper bound in the performance improvement as read depth increases (Supplementary Material Figure 1). While WaveQTL also shows improved performance for increased read depths, the rate of increase is relatively small compared to multiseq (Fig~\ref{fig:compareAUC}C).

\subsection{ATAC-seq data analysis}
% \subsection{Analysis of ATAC-seq data with small sample size 6 to identify regions with differential chromatin accessibility between copper-treated samples and control samples}
% ATAC-seq data \citep{buenrostro2013}, which measures the accessibility of chromatin along the genome, to identify regions with differential chromatin accessibility between two conditions. We chose a data set with small sample size 6 and compared multiseq with WaveQTL and DESeq2.

\subsubsection{Data} To understand regulatory mechanisms underlying cellular response to environmental perturbations, we collected ATAC-seq data 
to measure chromatin accessibility in samples treated with copper,
as well as two types of controls (control 1 and control 2). We
measured three samples for each condition, giving nine samples in all,
and generated a total of $\approx$ 627 M 38bp paired-end reads. See Supplementary Material for a detailed description of the data and preprocessing steps.  Our goal is to detect differences in chromatin
accessibility in copper-treated samples vs control. 
Since no difference in chromatin accessibility is expected between the two controls, we use these to 
construct an empirical null distribution of the test statistic for each method, and then compare the copper-treated sample to one of the controls. 

\subsubsection{Analysis} Accessible chromatin regions, which tend to contain functional elements of the genome, have a median length of about 300bp \citep{Degner_2012}. Thus, we analysed regions of length 1024bp, which are large enough to cover potential differences due to functional elements. (See \citet{Shim2015} for a discussion of robustness of multi-scale methods to choice of region size, and trade-offs between power and computation.) We focused our differential analysis on 242,714 regions that are the top 5\% of regions with the highest chromatin accessibility (see Supplementary Material for details). 

For each region, we applied multiseq, WaveQTL, and DESeq2 to the copper treated vs. control 1 samples, and computed the test statistic for each method for $H_0:$ no difference in chromatin accessibility between the conditions for the entire region. We used the likelihood ratio test statistic in~\eqref{eq:likeli_statistic} for multiseq and a similar likelihood ratio statistic for WaveQTL (see \citet{Shim2015} for details). We applied DESeq2 using the total read count over each region as input (bin size = 1024) and used a $p$-value for each region as a test statistic. We then assessed the significance of the test statistic for each method using its empirical distribution under $H_0$, constructed by applying the method to the two controls (control 1 vs. control 2). By this procedure, we obtained
for each method a $p$-value testing $H_0$ for each region
 which we converted to a $q$-value using the qvalue package \citep{qvalue_package}. We compared the methods by the number of differentially expressed regions (DERs) detected at a given $q$-value threshold (more DERs being better).

\subsubsection{multiseq outperforms WaveQTL} Fig~\ref{fig:compareMethods}A shows the number of DERs for each method as the False Discovery Rate (FDR) varies from 0.001 to 0.1. multiseq detects considerably more DERs than WaveQTL at all values of the FDR (the blue line for multiseq is above the orange line for WaveQTL).  For example, at FDR = 0.05, multiseq identifies 1083 DERs, which is 2.64 times the 411 DERs identified by WaveQTL. In addition, multiseq identifies most DERs found by WaveQTL (Fig~\ref{fig:compareMethods}B). 

\subsubsection{The advantages of multiseq are greatest for low-count regions}
\begin{figure}
\centering
\begin{subfigure}[b]{0.4\textwidth}
\centering
    {{\small A}}  
    \includegraphics[scale=0.11]{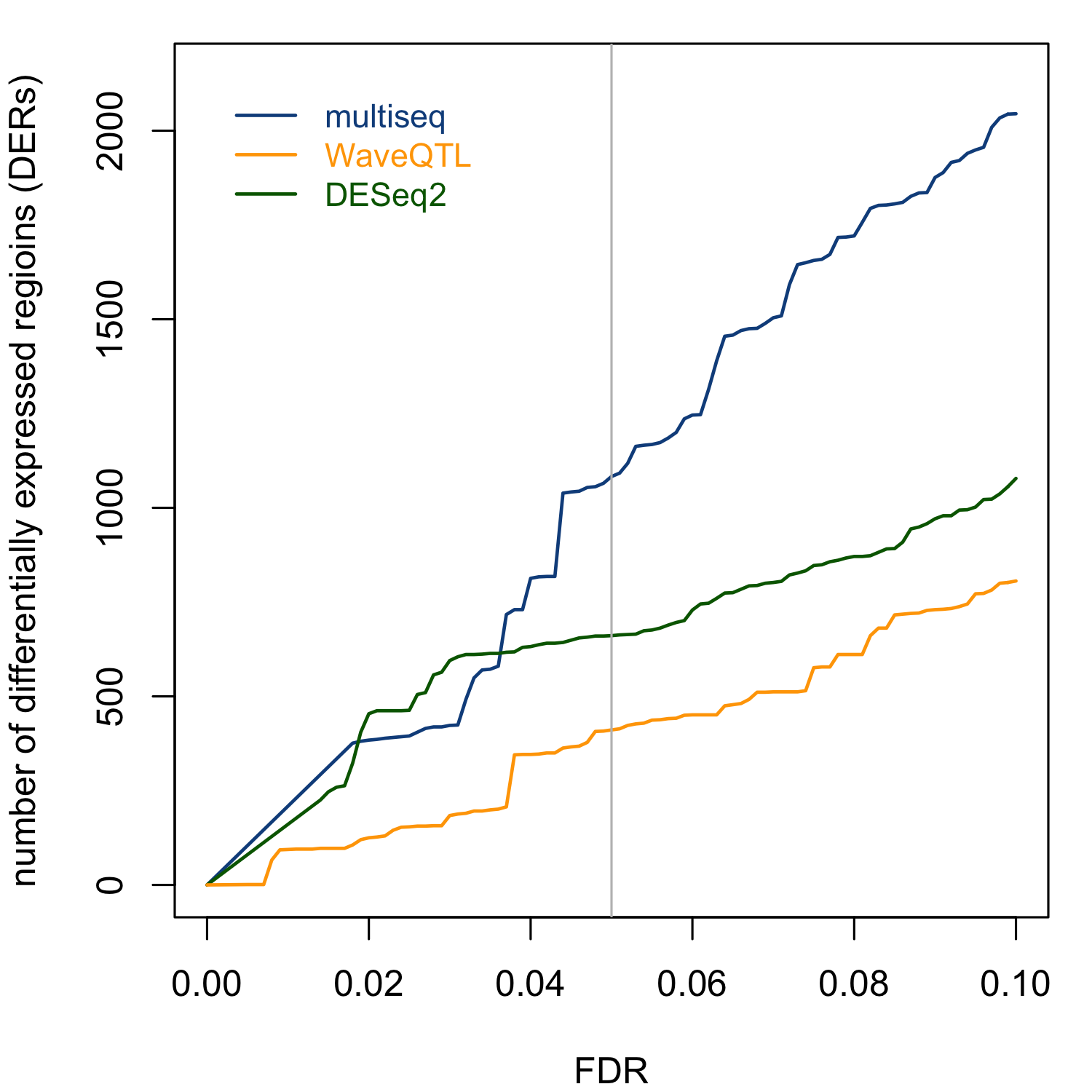} 
    %{{\small A}}    
    \label{fig:compareMethodsA}
\end{subfigure}
\hfill
\begin{subfigure}[b]{0.4\textwidth}
\centering
    {{\small B}}    
    \includegraphics[scale=0.11]{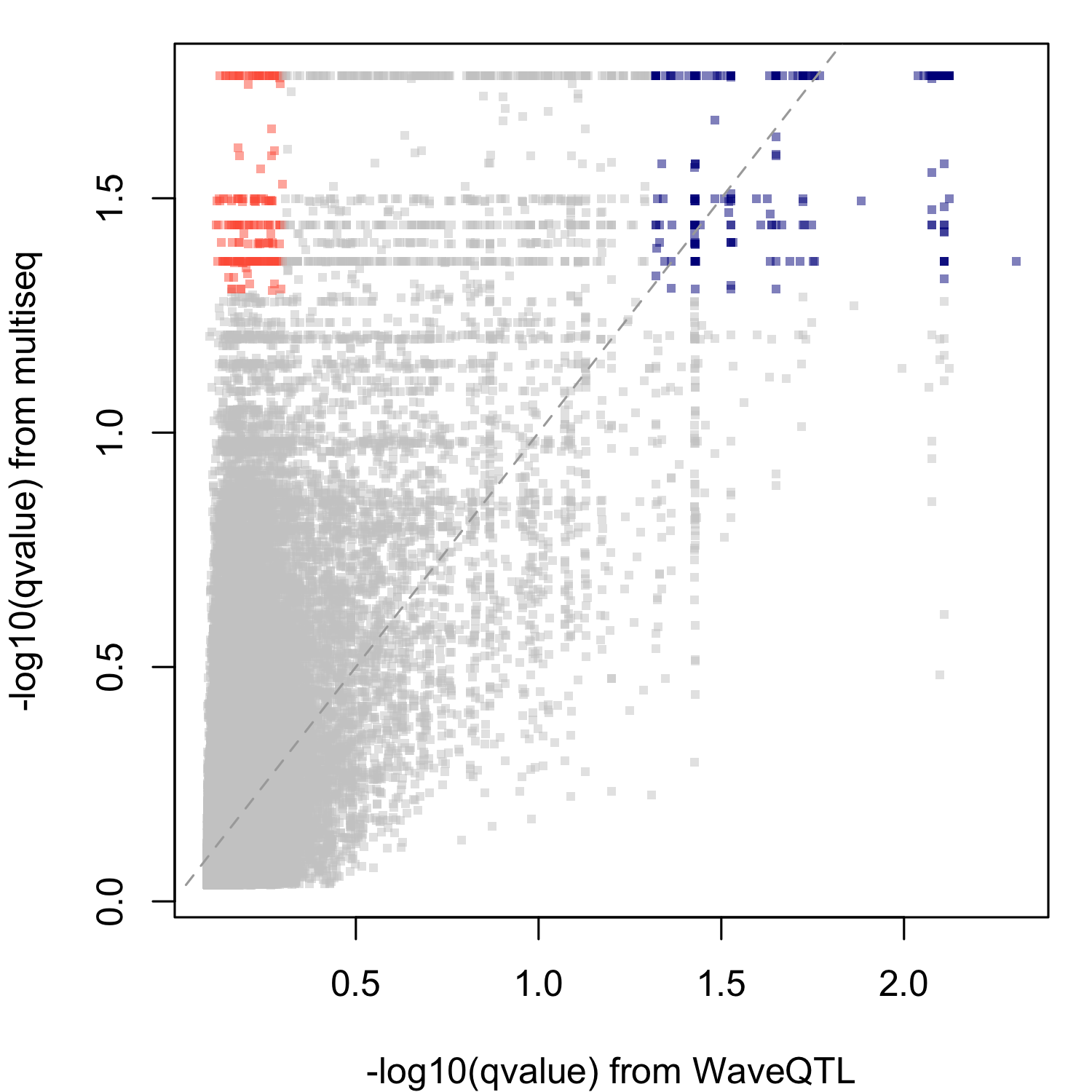}
   % {{\small B}}    
    \label{fig:compareMethodsB}
\end{subfigure}
\vskip\baselineskip
\begin{subfigure}[b]{0.4\textwidth}
\centering
    {{\small C}}   
    \includegraphics[scale=0.11]{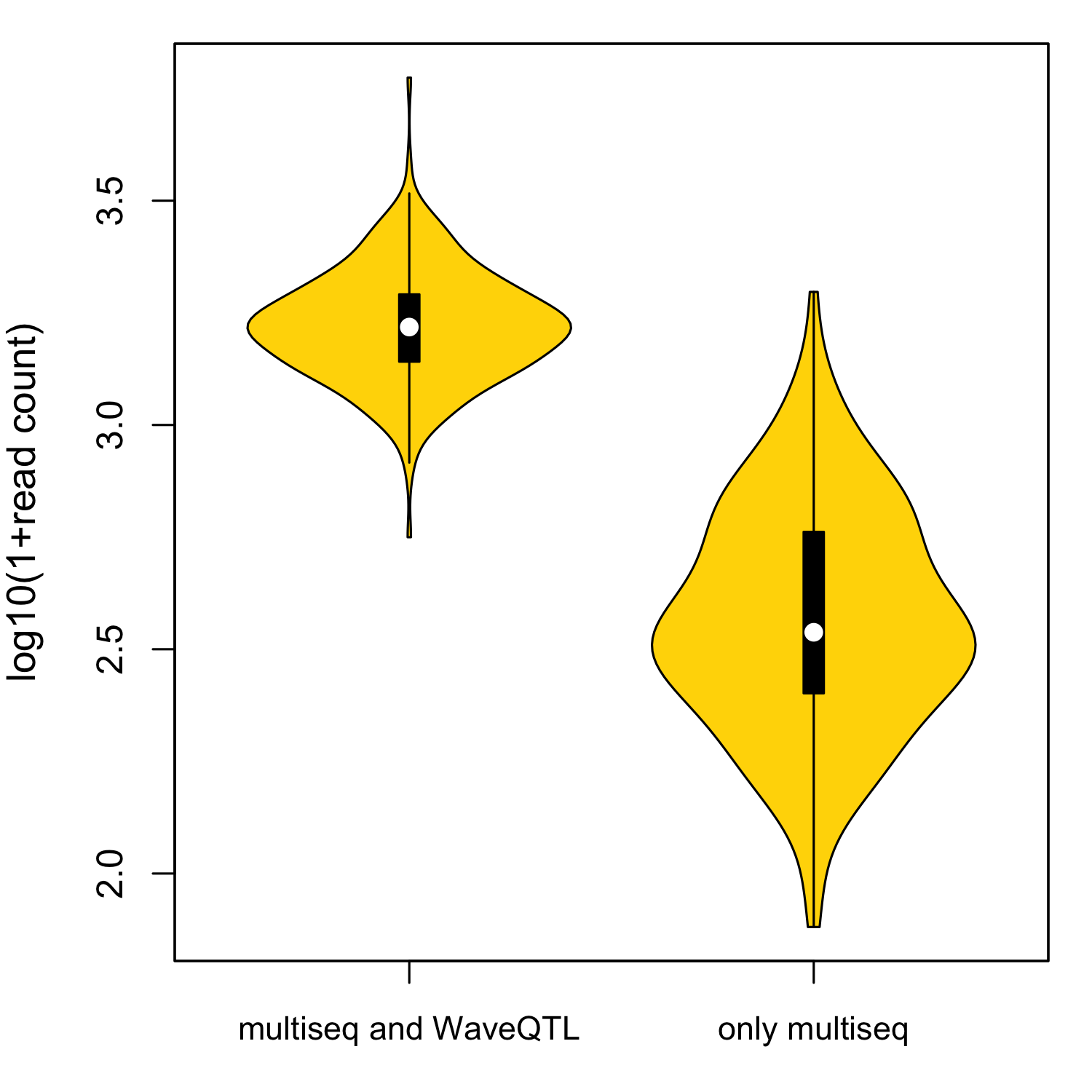} 
    %{{\small C}}    
    \label{fig:compareMethodsC}
\end{subfigure}
\hfill
\begin{subfigure}[b]{0.4\textwidth}
\centering
    {{\small D}} 
    \includegraphics[scale=0.11]{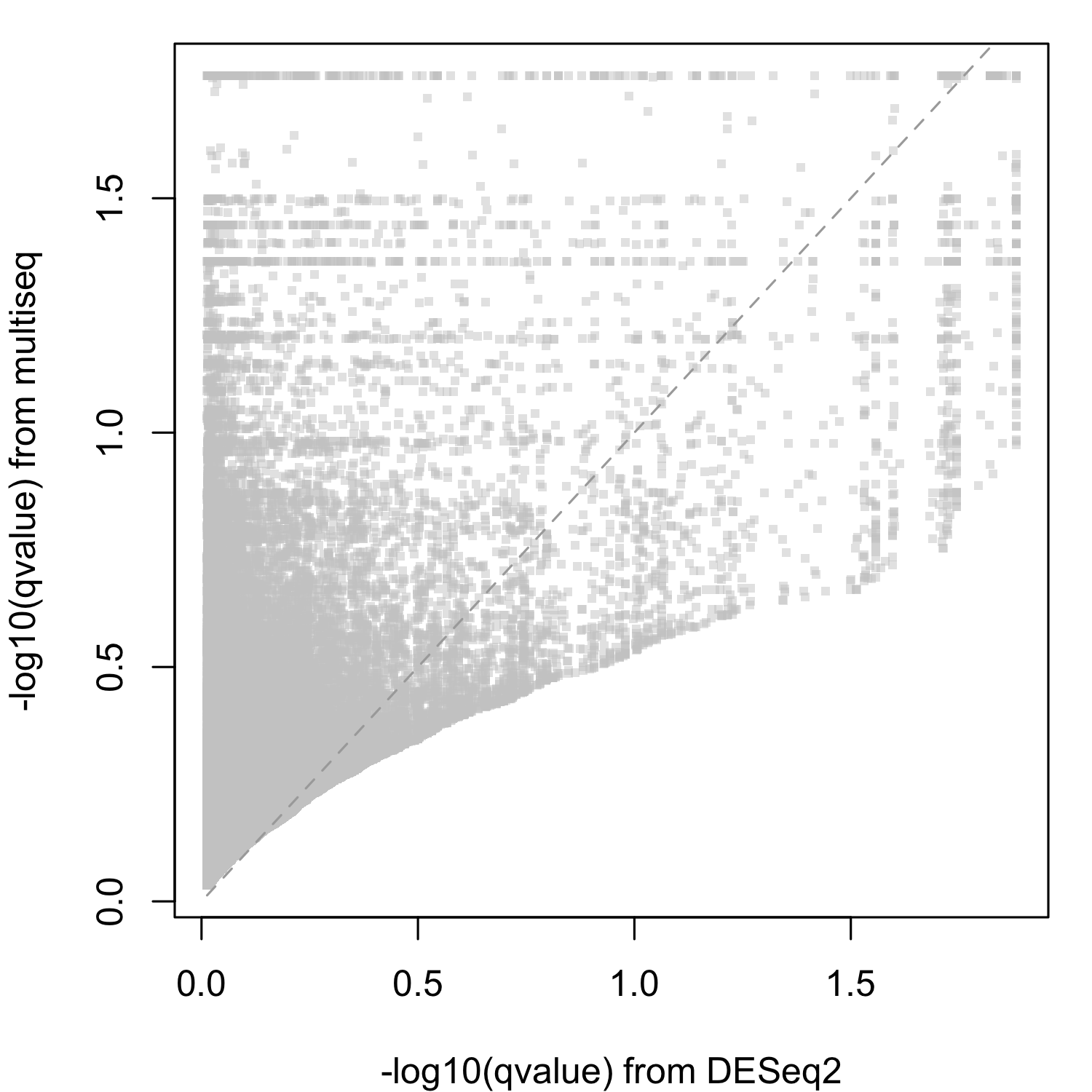}
    %{{\small D}}    
    \label{fig:compareMethodsD}
\end{subfigure}
\caption {{\bf Comparison of multiseq with WaveQTL and DESeq2.} Panel A shows the number of DERs identified by each method at a given FDR. The grey line indicates FDR = 0.05. Panel B shows a scatter plot of the $q$-values from multiseq versus the $q$-values from WaveQTL. The $q$-values are computed by the qvalue package \citep{qvalue_package}. The dashed line indicates the $y=x$ line. The 271 DERs detected by both methods at FDR = 0.05 are colored in blue. The 186 DERs detected by multiseq at FDR = 0.05, but missed by WaveQTL at FDR = 0.5, are colored in red. Most points are above or around the $y=x$ line, reflecting that most DERs identified by WaveQTL are detected by multiseq. Panel C shows the distributions of the total read count for two sets of regions: 1) DERs detected by both multiseq and WaveQTL (blue points in panel A) and 2) DERs detected by multiseq, but not WaveQTL (red points in panel A). For better visualization, regions with very large read counts (the 7 regions in set 1 with total read count $>$ 6000) are excluded from the distributions. Panel D shows a scatter plot of the $q$-values from multiseq versus the $q$-values from DESeq2. Most points are above or around the $y=x$ line, indicating that multiseq identifies most DERs detected by DESeq2.}
\label{fig:compareMethods}
\end{figure}

We hypothesized that the advantages of multiseq will be greatest
in regions of lower read counts. To assess this we compute the total read count across 6 samples for each region, and compared the distributions of these total read counts for two sets of regions: 1) 271 DERs detected by multiseq and WaveQTL (FDR = 0.05 for both methods, blue points in Fig~\ref{fig:compareMethods}B) and 2) 186 DERs detected by multiseq (FDR = 0.05), but not WaveQTL (FDR = 0.5, red points in Fig~\ref{fig:compareMethods}B). The results (Fig~\ref{fig:compareMethods}C) show that the DERs identified only by multiseq (set 2) have much lower read counts than those detected by both methods (median total read count: 1660 and 344 for the set 1 and set 2, respectively). This confirms
that multiseq is better powered for regions with lower read counts compared to WaveQTL. 
%Moreover, \textcolor{blue}{S2 Fig} shows that the total read counts of DERs identified by multiseq (set 1 and set 2) span a wide range of the read counts over 237K regions included in our analysis.
%To explore the potential of multiseq to identify DERs with lower read counts, we compute total read count across 6 samples for each region, and compare distributions of the total read count for three sets of regions: set 1) 271 DERs detected by multiseq and WaveQTL (FDR = 0.05 for both methods, blue points in Fig~\ref{fig:compareMethods}~B), set 2) 186 DERs detected by multiseq (FDR = 0.05), but not WaveQTL (FDR = 0.5, red points in Fig~\ref{fig:compareMethods}~B), and set 3) all 237K regions included in our analysis. Results (Fig~\ref{fig:compareMethods}~C) show that the DERs identified only by multiseq (set 2) have much lower read counts than those detected by both methods (median total read count: 1660 and 344 for the set 1 and set 2, respectively), which reflects that multiseq is better powered for regions with lower read counts compared to WaveQTL. Moreover, Fig~\ref{fig:compareMethods}~C shows that the read counts of DERs identified by multiseq (set 1 and set 2) span a wide range of read counts over the all 237K regions. \textcolor{red}{[HJ's comment: probably show the count distribution for identified by multiseq regardless of WaveQTL in supp?]}

\subsubsection{multiseq increases power using high-resolution information in the data, in addition to total read counts over the regions} 
A potential advantage of multiseq over overall expression methods (e.g., DESeq2 \citep{love2014DESeq2}, edgeR \citep{robinson2010edger}) is that
it can exploit the high-resolution information in the measurements. To assess the contribution of this feature to this analysis, we compared multiseq to DESeq2. Since multiseq uses the output from DESeq2 to compute a likelihood ratio measuring the support for the difference in overall expression (see Supplementary Material) and then combines it with the higher-resolution
information from its multi-scale model, any difference in performance between multiseq and DESeq2 is due to the additional information in the higher-resolution information. Fig~\ref{fig:compareMethods}A shows that multiseq increases power compared to DESeq2 for most values of the FDR, with slightly decreased power at FDR = 0.02 $\sim$ 0.036. The increased power indicates the contribution of the high-resolution information to the identification of regions with differential chromatin accessibility. Moreover, multiseq identifies the majority of DERs detected by DESeq2 (Fig~\ref{fig:compareMethods}D). See Supplementary Material for a comparison to DESeq2 with different bin sizes, where multiseq consistently outperforms them. 

\begin{figure}
\includegraphics[scale=0.068]{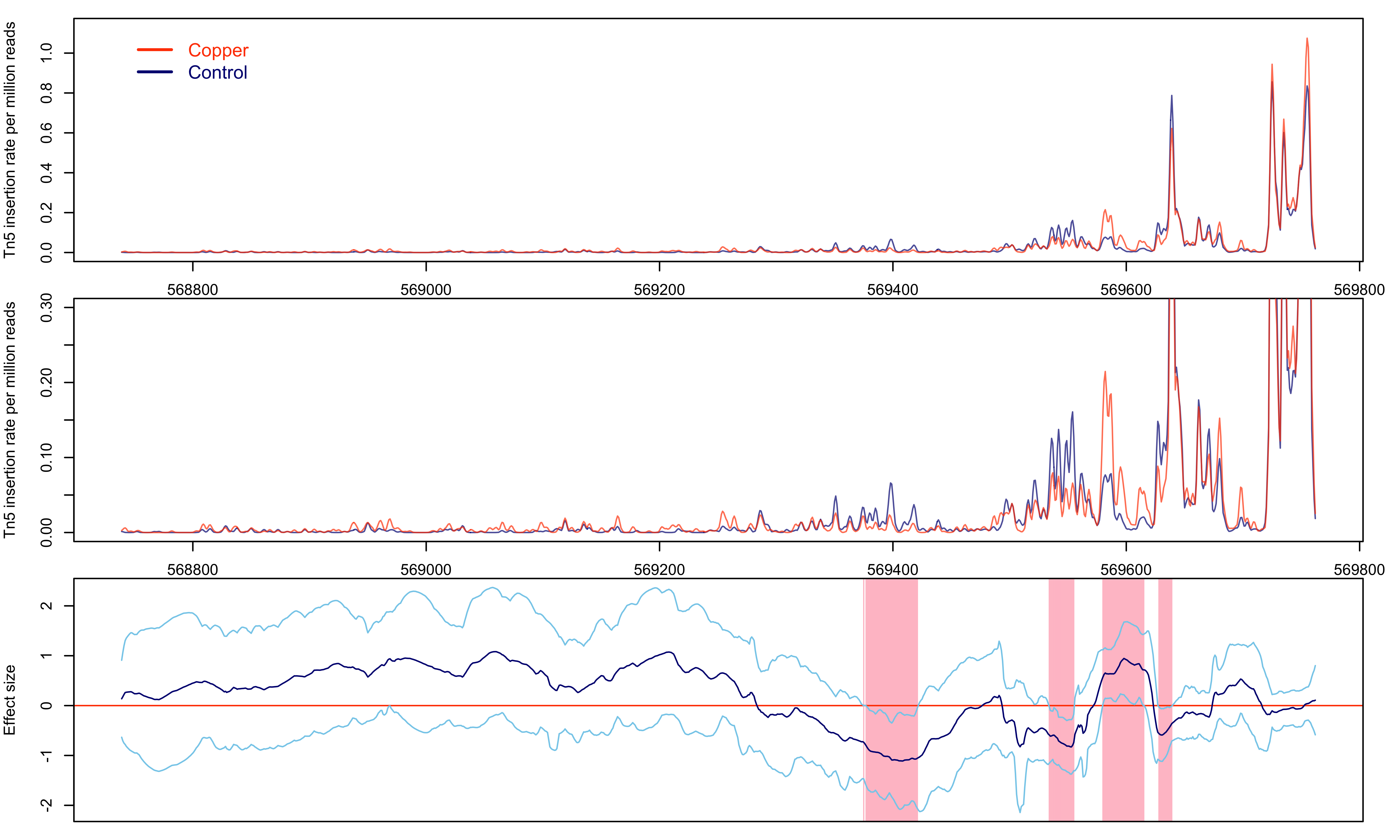}
\caption {{\bf Example of DERs identified by multiseq, but missed by DESeq2 (chr1:568739-569762).} The top two panels show the average Tn5 insertion rates along the region for each group with different limits of the y-axis. The bottom panel shows the posterior mean for effect size, i.e., the difference in log-intensity between the two groups (blue), and $\pm2.6$ posterior standard deviations (sky blue). Areas showing strong difference (zero is outside of the interval constructed by the two sky blue lines) are colored in pink. For multiseq, $\log\Lambda \approx 98.91$; $p$-value $\approx 0.0000034$. For DESeq2, $p$-value $\approx 0.19$.} 
\label{fig:exampleDER1}
\end{figure}

\begin{figure}
\includegraphics[scale=0.045]{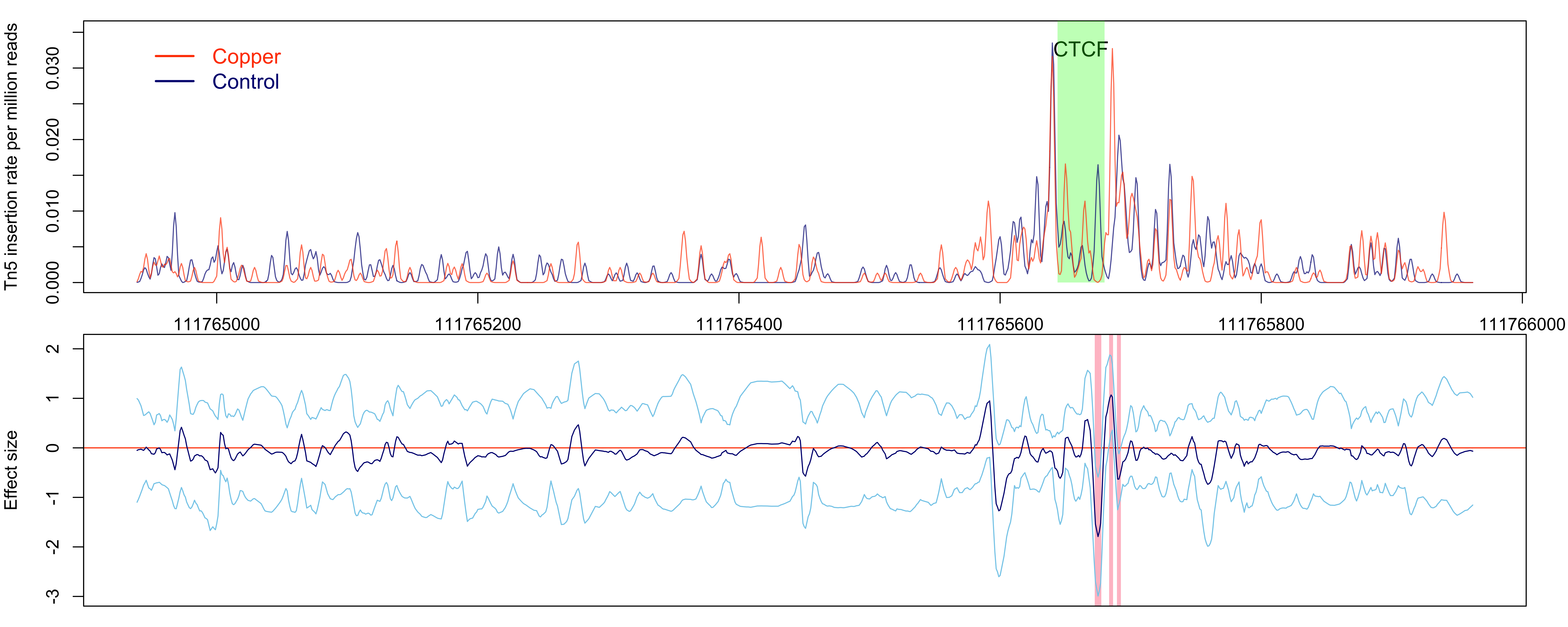}
\caption {{\bf Example of DERs identified by multiseq, but missed by DESeq2 (chr1:111764939-111765962).} Labels and colors are as in Fig~\ref{fig:exampleDER1}. In the top panel, the green block indicates the putative CTCF binding site, identified by the software CENTIPEDE \citep{Pique-regi_2011}. The bottom panel shows the posterior mean for effect size and $\pm1.5$ posterior standard deviations. For multiseq, $\log\Lambda = 10.49$; $p$-value $\approx 0.000019$. For DESeq2, $p$-value $\approx 0.23$.} 
\label{fig:exampleDER2}
\end{figure}

It is natural to ask what kinds of patterns of effects are better suited to multi-scale analysis.  \citet{Shim2015} showed that overall expression methods are well powered to identify effects appearing on most regions in the same direction, but that multi-scale methods have an advantage for effects over much shorter scales or in opposite directions. We observed consistent patterns in this study. Fig~\ref{fig:exampleDER1} and Fig~\ref{fig:exampleDER2} show two example of DERs identified by multiseq, but not DESeq2, which we now discuss in turn. 

The DER in Fig~\ref{fig:exampleDER1} shows strong effects in multiple areas. The effect in the first pink area is consistent in direction over about 50bp and the effects in the other pink areas are in opposite directions over about 100bp. DESeq2 missed this signal because the effects in opposite directions partially cancel each other out, leading to a weak overall signal in the 1024bp region. multiseq makes better use of the whole signal and easily captured it. Furthermore, we applied DESeq2 with smaller bin sizes (300bp and 100bp), and only DESeq2 with 100bp bin successfully detects the signal (Supplementary Material Figure 2). 

The DER in Fig~\ref{fig:exampleDER2} has effects in a relatively narrow area (the pink areas span $\approx$ 20bp). While multiseq captured this signal, DESeq2 failed to detect it partly because the signal affects the area much smaller than DESeq2 bin size (20bp $\ll$ 1024 bp), and also because the effects in opposite directions partially cancel each other out. DESeq2 with smaller bin sizes still missed this signal (Supplementary Material Figure 3). 

See \citet{Shim2015} for more extensive comparisons and discussions of the benefits of multi-scale approaches vs overall expression methods.

\subsubsection{Potential mechanism underlying the difference in chromatin accessibility}  
%In Fig~\ref{fig:exampleDER}, the narrow area with strong effect occurs at the right side of a binding site for CTCF (CCCTC binding factor), along with a moderate effect in the same direction at the left side of the binding site. 
In Fig~\ref{fig:exampleDER2}, the pink areas with effects occur near a binding site for CTCF (CCCTC binding factor). 
This suggests that the observed changes in chromatin accessibility in this region may be related to changes in CTCF binding. 
%Indeed, independent analysis of copper-treated and control samples using the software CENTIPEDE \citep{Pique-regi_2011} shows a decrease in CTCF binding strength on copper-treated samples (posterior odds: 540 in copper-treatment and 405 in control samples \textcolor{red}{HJ: Should check with Roger and Francesca; direction does not match the posterior odds.}), consistent with the observed decrease in chromatin accessibility around the CTCF binding site in Fig~\ref{fig:exampleDER}. 
Indeed, independent analysis of copper-treated and control samples using the software CENTIPEDE \citep{Pique-regi_2011} shows a change in CTCF binding strength between two conditions (posterior odds: 540 in copper-treatment and 405 in control samples). 
Moreover, using RNA-seq data from \citet{moyerbrailean2015} measuring gene expression on the same samples, we tested for differential gene expression in nearby genes, and observed a 23\% decrease of gene expression of CHI3L2, a gene within 1 Kb of this DER. It therefore seems plausible that the increase in CTCF binding in copper-treated samples insulates CHI3L2, causing the decrease in gene expression by insulating its promoter. (However, note that
this discussion of potential mechanism is inevitably speculative given the limits of current data.)

%\paragraph{multiseq also identifies differences in shape of chromatin accessibility.}
%\textcolor{red}{[HJ: I am not sure if this section is really necessary. Is there any motivating example where differential shape is of interests?]}
%$\hat\Lambda_{\text{shape}}$ (see equation (\ref{eq:LR.decompose}) in Methods section), the other component of $\hat\Lambda_{\text{ms}}$, measures the support in (standardized) shape for differential expression. Thus, when one is interested in differential shape analysis, it is natural to use $\hat\Lambda_{\text{shape}}$ as a test statistic (i.e., multiseq-shape). We applied multiseq-shape to the data set, and computed the number of DERs at a given FDR. The results (Fig~\ref{fig:compareMethods}~A) show that this data set contains moderate evidence in shape for differential expression, with much weaker signals than in total read counts (a dotted blue line for multiseq-shape is below a dashed blue line for multiseq-OE).

\subsubsection{Computation}
Analysis of the entire data set (242,714 tests of copper treatment vs control 1 and 242,714 tests of two controls) took about 103 CPU hours (user + system). This consisted of 10 hours for the data preprocessing and 93 hours for running multiseq. Because the analysis of each region is independent, the entire analysis is massively parallelizable in a naive way (on average 0.76 sec CPU time for each region). Software and scripts implementing our methods and analyses are available at \url{https://github.com/heejungshim/multiseq} and \url{https://github.com/heejungshim/multiseq_Shim_et_al}.
\section{Discussion}
We have developed a novel multi-scale method, multiseq, to estimate and test for differences in molecular phenotypes between multiple groups of samples using high-throughput sequencing data. The method is built on multi-scale models for inhomogeneous Poisson processes which enable multiseq to have two features: 1) better use of the high-resolution information in the data, and 2) direct modeling of the count nature of the data. The first feature allows multiseq to effectively detect differences that vary in scales or directions. Previous work \citep{Shim2015} demonstrated the advantages of multi-scale approaches over overall expression methods, and our experiments
support these results. The second feature, which is a main contribution of our work, allows multiseq to better maintain power in the analysis of data with small sample sizes or low read counts compared to the previous normal-based methods. We demonstrate this advantage on the simulation study with data sets of different sample sizes and library read depths, and the analysis of ATAC-seq data with small sample size 6. Finally, multiseq is computationally tractable for a large-scale differential expression analysis involving hundreds of thousands of tests.

Although multiseq was motivated by differential high-throughput sequencing data analysis for molecular phenotypes, it could be more generally applied to association analysis between a sequence of counts and other covariate, either continuous or discrete. For example, it could be used to detect associations between a molecular phenotype, such as chromatin accessibility measured by the sequencing data, and a continuous covariate, such as gene expression. Or, it could be used to detect and estimate differences in 
the intensity of any two (nonhomogenous) Poisson processes unrelated to genomics, such as gamma-ray burst signals \citep{Kolaczyk_1999} in astronomy, between two conditions. 

In our ATAC-seq data analysis, we constructed the empirical null distribution of the test statistic using a {\it null} data set that is expected to have no differences in chromatin accessibility between conditions. This distribution is required to obtain $p$-values that are used to compute $q$-values and estimate FDR. Alternatively, a null distribution could be generated by permuting sample labels (e.g., see \citet{Shim2015}), or a Bayesian FDR could be used to address the multiple testing issue, as it does not require $p$-values. For example, \citet{morris2008} used wavelet-based functional mixed models to identify positions (or bins) with absolute values of effect sizes $> \delta$, and introduced an approach for controlling the Bayesian FDR that uses the posterior distributions of effect sizes. \citet{ma2018} also controlled the Bayesian FDR to identify resolutions/locations with significant signal in their approach for analysis of distributional variance.

We have demonstrated that multiseq outperforms a wavelet-based approach, particularly in the analysis of data with small sample sizes or low read counts. However, there are still opportunities for potential improvements. First, the normal approximations to binomial likelihoods use the estimates of parameters and their standard errors, but with small sample sizes, the estimated standard errors  can be less stable. In genomics, this issue has been addressed by using the shrinkage of the estimated sample variances to pooled estimates which are more stable \citep{smyth2004}. Incorporating the shrinkage estimates to multiseq could potentially further improve performance in small sample sizes. Second, our model assumes independent priors on effect sizes across scales and locations, but in practice the strengths of effects in multi-scale models tend to have dependencies - they tend to propagate across adjacent locations and scales \citep{Crouse1998}. Priors that exploit the dependencies, such as the tree-like structure described in \citet{Crouse1998}, can effectively combine information across different scales and locations. Extending multiseq to impose such priors could further improve performance.

%%%%%%%%%%%%%%%%%%%%%%%%%%%%%%%%%%%%%%%%%%%%%%
%% Single Appendix:                         %%
%%%%%%%%%%%%%%%%%%%%%%%%%%%%%%%%%%%%%%%%%%%%%%
%\begin{appendix}
%\section*{???}%% if no title is needed, leave empty \section*{}.
%\end{appendix}
%%%%%%%%%%%%%%%%%%%%%%%%%%%%%%%%%%%%%%%%%%%%%%
%% Multiple Appendixes:                     %%
%%%%%%%%%%%%%%%%%%%%%%%%%%%%%%%%%%%%%%%%%%%%%%
\section*{Supplementary material}
Supplementary material is available at \url{https://github.com/heejungshim/multiseq_Shim_et_al/blob/main/manuscript/multiseq_supp.pdf}.

%%%%%%%%%%%%%%%%%%%%%%%%%%%%%%%%%%%%%%%%%%%%%%
%% Support information (funding), if any,   %%
%% should be provided in the                %%
%% Acknowledgements section.                %%
%%%%%%%%%%%%%%%%%%%%%%%%%%%%%%%%%%%%%%%%%%%%%%
\section*{Acknowledgements}
We thank Jack Degner for invaluable discussions and Yao-ban Chan for helpful comments on a draft manuscript. We also thank the members of the H. Shim, M. Stephens, and J. Pritchard labs for helpful discussions. This work was supported by NIH grant HG002585.

% The authors would like to thank ...
% 
% The first author was supported by ...
% 
% The second author was supported in part by ...
 
%%%%%%%%%%%%%%%%%%%%%%%%%%%%%%%%%%%%%%%%%%%%%%
%% Supplementary Material, if any, should   %%
%% be provided in {supplement} environment  %%
%% with title and short description.        %%
%%%%%%%%%%%%%%%%%%%%%%%%%%%%%%%%%%%%%%%%%%%%%%
%\begin{supplement}
%\stitle{???}
%\sdescription{???.}
%\end{supplement}

%%%%%%%%%%%%%%%%%%%%%%%%%%%%%%%%%%%%%%%%%%%%%%%%%%%%%%%%%%%%%
%%                  The Bibliography                       %%
%%                                                         %%
%%  imsart-nameyear.bst  will be used to                   %%
%%  create a .BBL file for submission.                     %%
%%                                                         %%
%%  Note that the displayed Bibliography will not          %%
%%  necessarily be rendered by Latex exactly as specified  %%
%%  in the online Instructions for Authors.                %%
%%                                                         %%
%%  MR numbers will be added by VTeX.                      %%
%%                                                         %%
%%  Use \cite{...} to cite references in text.             %%
%%                                                         %%
%%%%%%%%%%%%%%%%%%%%%%%%%%%%%%%%%%%%%%%%%%%%%%%%%%%%%%%%%%%%%

%% if your bibliography is in bibtex format, uncomment commands:
\bibliographystyle{imsart-nameyear} % Style BST file
%\bibliography{bibliography}       % Bibliography file (usually '*.bib')
\bibliography{multiseq}

%% or include bibliography directly:
% \begin{thebibliography}{}
% \bibitem[\protect\citeauthoryear{???}{???}]{b1}
% \end{thebibliography}

\end{document}